# First-Order Spin-Reorientation Transition and Incomplete Softening of the Antiferromagnetic Resonance Mode in Multiferroic GdFe$_3$(BO$_3$)$_4$

I.N. Khoroshiy[1,2], S.A. Skorobogatov[1,2], S.E. Nikitin[1], I.A. Gudim[1], V.R. Titova[1], A.I. Pankrats[1,2]

[1] *Kirensky Institute of Physics, Krasnoyarsk Scientific Center, Siberian Branch, Russian Academy of Sciences, Krasnoyarsk, 660036 Russia*
[2] *Siberian Federal University, Krasnoyarsk, 660041 Russia*

**Abstract**

The multiferroic ferroborate GdFe$_3$(BO$_3$)$_4$ with huntite-type structure exhibits magnetic ordering below $T_N$ = 38 K and contains two magnetic subsystems associated with Gd and Fe ions. Competing anisotropies of these subsystems drive a spin-reorientation transition at $T_{SR}$ = 10.7 K, switching the ground state from easy-axis to easy-plane. Using antiferromagnetic resonance, we investigate the spin dynamics across this transition. The observed incomplete softening of a magnon mode during both field- and temperature-induced spin-reorientation transitions indicates the first-order nature of the phase transition, which is accompanied by a discontinuous jump in the effective anisotropy field. We reproduce this behavior using a simple model that attributes the jump in the anisotropy field to the presence of an effective fourth-order anisotropy constant, responsible for the discontinuous character of the transition. Remarkably, for in-plane magnetic fields, we identify a new AFMR mode that persists from 12 K up to $T_N$. This mode likely corresponds to the dynamics of a long-period incommensurate state, previously detected by resonant elastic X-ray scattering.

1. **Introduction**

In the last couple of decades, there has been a keen interest in studying the physical properties of rare-earth borates RM$_3$(BO$_3$)$_4$ (rare-earth ions R = Y, La–Lu; M = Fe, Cr, Al, Ga) with a huntite structure crystallizing in noncentrosymmetric sp. gr. *R*32 or *P*3$_1$21. This is due to their nonlinear-optical [1–3] and multiferroic [4–5] properties and also the rich spectrum of the magnetic structures observed in this class of compounds. Depending on the sort of the R$^{3+}$ ion, rare-earth borate crystals can exhibit the properties of simple antiferromagnets with the antiferromagnetic axis oriented either along the crystal trigonal axis (easy anisotropy axis (EA) at R = Tb, Dy, Pr [6–8]) or perpendicular to it in the basal plane (easy anisotropy plane (EP) at R = Sm, Er, Eu [5, 8]). For certain R ions, the elastic neutron scattering and X-ray resonant magnetic scattering studies showed the weak noncollinearity (R = Er [8] and Ho [9]) or the incommensurate magnetic structure (R = Nd [10] and Gd above 10 K [11]).

The variety of magnetic structures of the discussed compounds is related to the coexistence of two magnetic subsystems – Fe$^{3+}$ and R$^{3+}$ ions – in the crystal. At Néel temperature T$_N$, the iron subsystem is ordered antiferromagnetically and exhibits the EP magnetic anisotropy. The R ion subsystem, in which the intrinsic exchange coupling is weak, is ordered antiferromagnetically due to the f–d exchange interaction with the iron subsystem. As a result of the f–d interaction, the R subsystem contributes to the total effective anisotropy of the crystal. Depending on the sign and value of the magnetic anisotropy constant of the R ion, this contribution either enhances the EP anisotropy of the iron subsystem or competes with it, providing a wide range of possible magnetic structures.

The most interesting situation, promising for studying orientational magnetic transitions in these crystals appears when the competing $R^{3+}$ and $Fe^{3+}$ contributions have similar absolute values. In this case, the effective magnetic anisotropy of the crystal lowers, and the difference between the temperature dependences of the subsystem contributions can lead to the mutual compensation of the latter at a certain temperature and to a change in the total magnetic anisotropy sign, thereby inducing a spontaneous EA–EP spin reorientation (SR). In addition, the magnetic field dependence of the magnetic anisotropy of the $R^{3+}$ ion can evoke an SR not only upon temperature variation, but also due to application of a magnetic field. This is exactly the situation with the competition of contributions of the $R^{3+}$ ions and the $Fe^{3+}$ subsystem found in gadolinium [12, 13] and holmium ferroborates [9, 14].

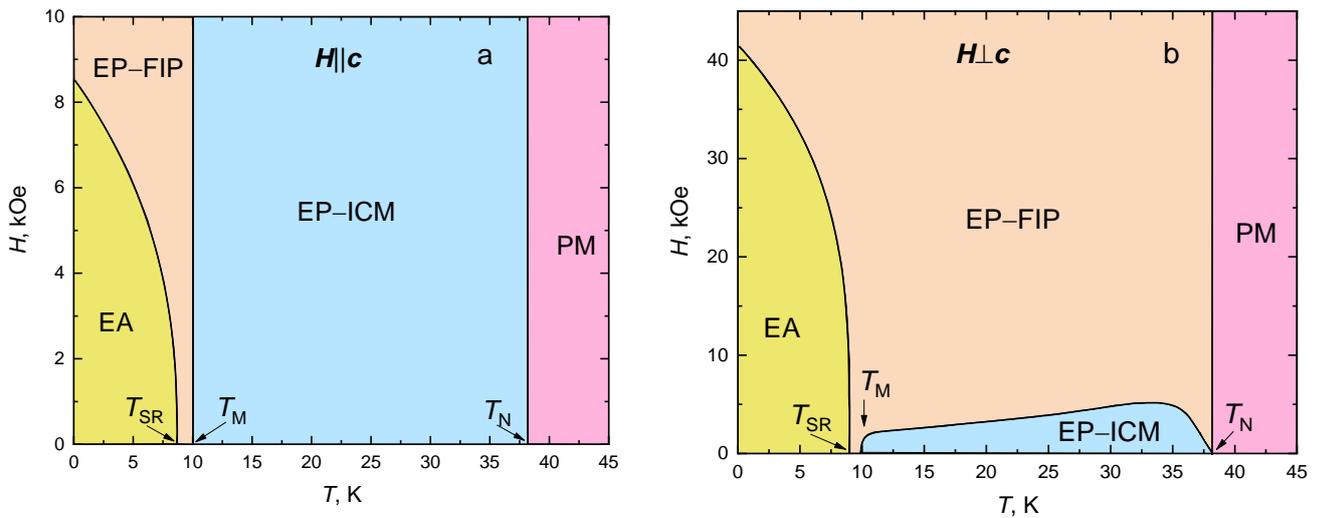

Fig. 1. Magnetic phase diagrams for two field orientations $H\|c$ (a) and $H\perp c$ (b). States: EA – easy anisotropy axis, EP–ICM – incommensurate easy anisotropy plane, EP–FIP – commensurate easy anisotropy plane with field-induced polarization, PM – paramagnetic. Note that these diagrams were constructed using the data from [11,16,17,35] for samples with Bi impurities and therefore the phase boundaries are slightly shifted compare to the pure sample studied in this work.

Gadolinium ferroborate $GdFe_3(BO_3)_4$ was the first ferroborate-group crystal the magnetic structure of which was examined by antiferromagnetic resonance (AFMR) spectroscopy [12, 13]. This compound crystallizes in sp. gr. $R32$ and, at $T = 155$ K, as a result of the structural transition, its symmetry reduces to sp. gr. $P3_121$ [15]. AFMR data indicate that below the Néel temperature $T_N = 38$ K, the EP antiferromagnetic ordering is established in the crystal and magnetic moments of $Fe^{3+}$ and $Gd^{3+}$ ion lye in its basal plane [12, 13]. Upon further cooling, a spontaneous SR from the EP to EA antiferromagnetic structure occurs at $T_{SR} \approx 8.5$ K[1]. This transition is caused by a change in the total magnetic anisotropy sign. In [13], the resonance data obtained in magnetic fields parallel and perpendicular to the crystal trigonal axis were used to construct magnetic phase diagram of $GdFe_3(BO_3)_4$. The established phase boundaries were later confirmed by studies of the magnetostriction and dielectric properties [5, 16, 17]. Further resonant elastic X-ray scattering (REXS) investigations of the magnetic structure [11] supplemented the phase diagram with an incommensurate magnetic (ICM) state that occurs in the EP state at temperature $T_M$ higher than $T_{SR}$ by ~2 K. It was shown that a magnetic field applied in the basal plane suppresses the incommensu-rate state and transfers the crystal to the commensurate phase, in which the field-induced polarization (FIP) occurs. At the same time, the magnetic field $\boldsymbol{H} \| \boldsymbol{c}$ does not affect the incommensurate state. We used this data to show the schematic phase diagram of $GdFe_3(BO_3)_4$ for two field directions in Figs. 1a and 1b.

---
[1] This value is correct for old Bi-contaminated samples

The two main states of the phase diagram of $GdFe_3(BO_3)_4$, EA and EP, can be fully attributed to the magnetic order of the iron subsystem, although the authors of [11, 18] showed that, in the temperature range of $T_{SR} < T < T_N$, the iron magnetic moments cant out of the basal plane at a certain angle. The gadolinium subsystem is ordered antiferromagnetically owing to the *f–d* interaction with the iron subsystem at temperatures $T < T_N$. However, there is no consensus regarding the orientation of the gadolinium moments in different parts of the phase diagram. In particular, the authors of [11] argued that both subsystems undergo an SR at $T = T_{SR}$, but in the FIP phase induced by the field $\boldsymbol{H} \perp \boldsymbol{c}$, the gadolinium moments make an angle of ~45° with the basal plane. At the same time, it should be taken into account that the resonance properties of the $GdFe_3(BO_3)_4$ compound examined in this work are determined precisely by the iron subsystem, since, due to the strong difference in the natural resonance frequencies of the magnetic subsystems, their oscillations can be considered uncoupled [19]. Therefore, the role of the gadolinium subsystem is reduced to an additional contribution to the total magnetic anisotropy of the crystal.

The R subsystem can play a special role in spin-reorientation transition. The AFMR spectroscopy study of the $Pr_{1-x}Y_xFe_3(BO_3)_4$ crystals [20] showed that, in pure ($x = 0$) and slightly diluted ($x = 0.25$) crystals, there is a very wide (up to ~17 kOe) lability range with the coexisting collinear and spin-flop states in the region of the spin-flop transition. It was established that the main reason for such a wide range is the significant difference between the magnetoanisotropic properties of the Pr subsystem in the collinear and spin-flop states.

It would be reasonable to assume that a similar effect accompanies spin-reorientation transitions in ferroborate crystals containing other R ions. The change in the magnetic anisotropy of the R ion during the transition can be minor, but, under the conditions of the almost complete mutual compensation of the contributions of the Fe and R subsystems, this effect can be pronounced. Therefore, it seems interesting to examine the temperature and field dependences of the effective magnetic anisotropy of the $GdFe_3(BO_3)_4$ crystals in the region of spin-reorientation occurring in the crystals upon temperature and magnetic field variation. In addition, it is important to investigate the resonance properties of the $GdFe_3(BO_3)_4$ crystals in the weakly explored magnetic phases at $T > T_{SR}$.

## 2. Experimental Details
### 2.1. Sample preparation

The resonance experiments were carried out on the $GdFe_3(BO_3)_4$ single crystals grown on a seed by solution–melt crystallization from a lithium tungstate-based solvent. The technique used for growing single crystals was described in detail in [21]. The obtained bulk transparent green crystals had the shape of a hexagonal prism with well-developed faces. In some crystals, the prism apex had a face with the shape of a regular triangle, the plane of which coincided with the basal plane. The crystals were oriented according to their habit. Two of the highest-quality single crystals with maximum sizes of ~1 and 2 mm were selected for the resonance study. Since the resonance properties of the two crystals were fully identical, the main measurements were performed on the smallest sample weighing 2.0 mg with a smooth single resonance line over the entire investigated frequency range. The resonance lines for the second sample weighing about 5 mg at frequencies above ~45 GHz contained additional peaks caused by the nonuniformity of the microwave field in the crystal. This sample was used in the measurements that required a higher sensitivity.

*2.2. Resonance and magnetic measurements*

The AFMR study of the GdFe$_3$(BO$_3$)$_4$ single crystals was carried out on a custom magnetic resonance spectrometer with a broad frequency band (25–140 GHz) in pulsed magnetic fields of up to 90 kOe [22]. Due to the pulsed nature of the magnetic field in AFMR measurements, we have the possibility to compare the resonant spectra recorded in ascending and descending magnetic fields on the leading and trailing edges of the pulse, respectively, which is important when studying hysteresis phenomena. To refine the phase boundary between ICM and FIP magnetic phases, the magnetic properties of the crystals were measured on a Quantum Design PPMS-9 Physical Property Measurement System.

### 3. Experimental Results

The magnetic phase diagrams in Figures 1a and 1b were constructed using various experimental techniques for GdFe$_3$(BO$_3$) crystals containing Bi impurities [11,16,17,35]. Our previous AFMR studies of this compound were also performed on such Bi-contaminated samples [12, 13]. In contrast, this work investigates a nominally pure GdFe$_3$(BO$_3$) crystal, synthesized via a different solvent method [21]. Our AFMR measurements reveal no qualitative differences between the pure and Bi-contaminated crystals. We conclude that bismuth impurities (up to 6 at.%) only slightly reduce the spin-reorientation temperature, $T_{SR}$, and the critical fields of the transition to the induced EP state at $T < T_{SR}$, for both **H**||**c** and **H**⊥**c** orientations. Specifically, $T_{SR}$ decreases from ~10.5 K in pure crystals to ~8.5 K in Bi-containing ones. A more detailed analysis of the impurity effect on the magnetic phase diagram is provided in Ref. [21].

*3.1. Magnetic field **H**//**c**.*

Figure 2 shows the AFMR frequency–field dependences measured at a temperature of $T = 4.2$ K in the magnetic field **H**||**c**. In these curves, two parts corresponding to the magnetic phase diagram in Fig. 1a can be distinguished. In the weak magnetic field region, two oscillation modes ($\nu_{\|1}^{EA}$ and $\nu_{\|2}^{EA}$) are observed, the frequencies of which depend linearly on the field and the linewidths do not exceed 300 Oe in the frequency range of 26 GHz. The field boundary of these regions is indicated in Fig. 2 by a narrow pink strip. The AFMR frequency–field dependences for modes $\nu_{\|1}^{EA}$ and $\nu_{\|2}^{EA}$ in this region are characteristic of the EA antiferromagnetic state [23] in fields below the spin-flop field:

$$\frac{\nu_{\|1,2}^{EA}}{\gamma_\|} = \left[(2H_E + H_A)H_A\right]^{1/2} \pm H, \tag{1}$$

where $\nu_{\|c}^{EA} = \gamma_\| H_\Delta = \gamma_\| \left[(2H_E + H_A)H_A\right]^{1/2}$ – is the zero-field spin gap.

Here $H_E$ and $H_A$ are effective fields of exchange interaction and magnetic anisotropy, respectively; $H_A > 0$ in the EA state. Solid lines in Fig. 2 correspond to theoretical dependence (1) with the parameters ($T = 4.2$ K): $\nu^{EA}_{\|c} = (31.0 \pm 0.1)$ GHz and $\gamma_\| = (2.60 \pm 0.05)$ GHz/kOe.

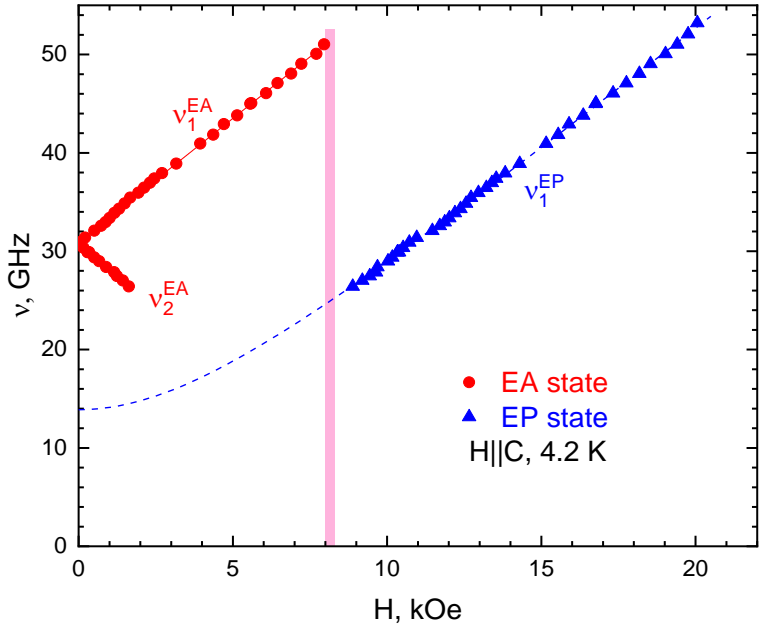
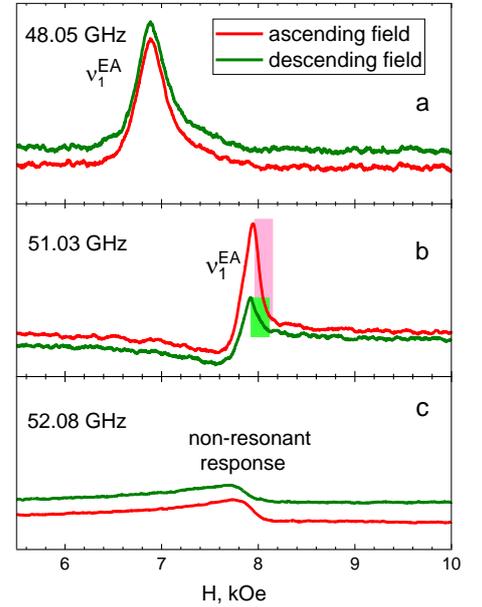

Fig. 2. AFMR frequency–field dependences measured at $T = 4.2$ K in magnetic field $\boldsymbol{H}\|\boldsymbol{c}$.

Fig. 3. AFMR spectra recorded in the EA phase at different frequencies. $T = 4.2$ K, $\boldsymbol{H}\|\boldsymbol{c}$.

The boundary of the EA phase is marked by the disappearance of the high-frequency resonance line of oscillation mode $\nu^{EA}_{\|1}$. Figure 3a shows a spectrum at 48.05 GHz, where the resonance field remains well within the EA phase. At this frequency, the resonance lines recorded for ascending and descending fields are identical. At 51.03 GHz (Fig. 3b), segments of the absorption lines vanish within specific field ranges (highlighted in pink for ascending and green for descending fields). A slight misalignment between these ranges for different field sweep directions results in a lower residual line intensity for the descending field sweep. This spectral evolution is caused by the field hysteresis of the first-order spin-reorientation transition. With a further frequency increase to 52.08 GHz, the resonant absorption for this mode vanishes completely, leaving only a weak anomaly arising from a non-resonant response of the microwave signal to a change in the magnetic state of the crystal.

In the field region above the boundary, there is resonance mode $\nu^{EP}_{\|1}$ with a frequency–field dependence characteristic of the EP antiferromagnetic state. At $T = 4.2$ K, this state is induced by a magnetic field. The AFMR frequency–field dependences for this state are [23]:

$$\left(\frac{\nu^{EP}_{\|1}}{\gamma_\|}\right)^2 = 2H_E|H_A| + H^2, \qquad (2)$$
$$\nu^{EP}_{\|2} = 0.$$

Here, $H_A < 0$ is the effective anisotropy field of the crystal in the EP state. The dotted line in Fig. 2 for oscillation mode $\nu_{\|1}^{EP}$ corresponds to this equation with the spin gap $\nu_{\|c}^{EP} = \gamma_\| \sqrt{2 H_E |H_A|} = (13.9 \pm 0.4)$ GHz. The second AFMR mode $\nu_{\|2}^{EP}$ in this state is gapless (the goldstone mode).

Figure 4 shows temperature dependences of the AFMR resonance fields measured at different frequencies. In these dependences, the resonance data related to the EA (modes $\nu_{\|1}^{EA}$ and $\nu_{\|2}^{EA}$) and EP (mode $\nu_{\|1}^{EP}$) states can be distinguished. In the EA state, when measured at all frequencies above the gap $\nu_{\|c}^{EA}$, the resonance field increases with temperature. This behavior is explained by a decrease in the gap due to a decrease in the effective anisotropy field of the crystal. The corresponding temperature dependence of the gap $\nu_{\|c}^{EA}$ calculated from the data for the EA state is presented in

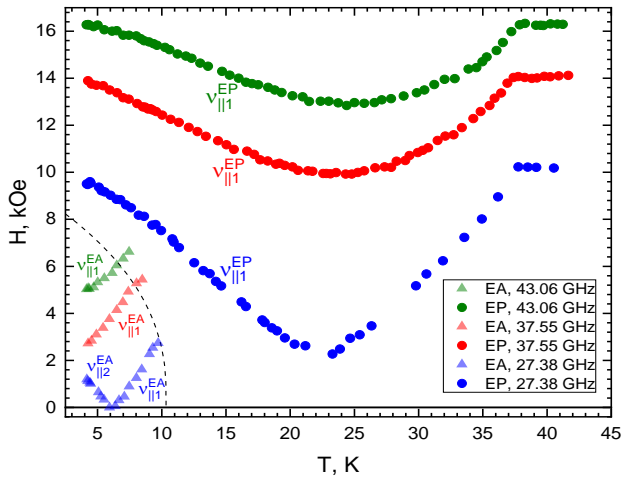 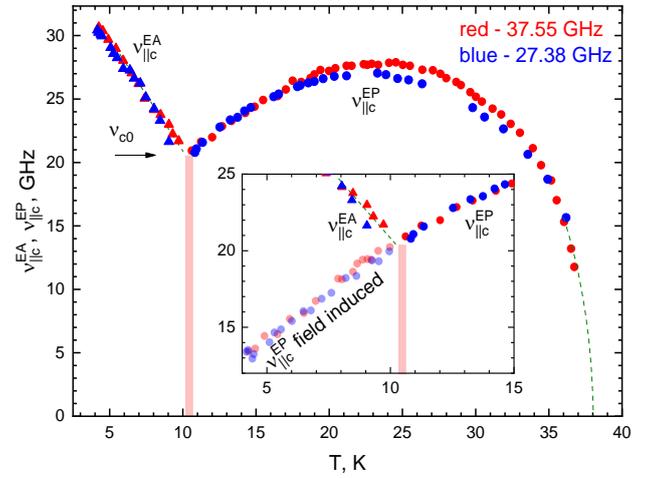

Fig. 4. Temperature dependences of the AFMR resonance fields at **H**||**c**. The dotted line indicates the boundary between the EA and EP states from [21].

Fig. 5. Temperature dependences of the spin gaps in the spectrum for the EA and EP states measured at frequencies of 27.38 and 37.55 GHz. The insert shows a fragment of the dependences with additional data for field induced EP state.

Fig. 5. The more complex dependence of the resonance field observed at 27.38 GHz – a frequency below the spin gap at T = 4.2 K – arises from a transition between oscillation modes. Specifically, as the temperature increases and the spin gap softens, the resonant response shifts from branch $\nu_{\|2}^{EA}$ to branch $\nu_{\|1}^{EA}$. For any given frequency, the maximum resonance fields observed within the EA state correspond to the critical fields of the EA–EP spin-reorientation transition.

In the EP state, the resonance fields initially decrease with rising temperature, reach a minimum at 20–25 K, and subsequently increase, approaching the paramagnetic state's resonance field values near the Néel temperature ($T_N$ = 38 K). The corresponding non-monotonic temperature dependences of the spin gap $\nu_{\|c}^{EP}$, measured at two frequencies, are also presented in Fig. 5. A pink band indicates the temperature range of the spontaneous EA–EP reorientation.

### 3.2. Magnetic field **H**⊥**c**

The resonance properties of the GdFe$_3$(BO$_3$)$_4$ single crystals in the magnetic field oriented in the crystal basal plane are also consistent with the magnetic phase diagram from Fig. 1b. The

AFMR frequency–field dependences for the GdFe$_3$(BO$_3$)$_4$ single crystal at this field orientation are shown in Fig. 6.

If the effective anisotropy is weak enough, $H_A \ll 2H_E$, which is the case in gadolinium ferroborate, the field–frequency dependences for the EA state have the form [23]

$$\left(\frac{\nu_{\perp 1}^{EA}}{\gamma_\perp}\right)^2 = (2H_E + H_A)H_A + \frac{2H_E - H_A}{2H_E + H_A}H^2 \approx (2H_E + H_A)H_A + H^2, \quad (3)$$
$$\left(\frac{\nu_{\perp 2}^{EA}}{\gamma_\perp}\right)^2 = (2H_E + H_A)H_A - \frac{H_A}{2H_E + H_A}H^2.$$

In Eq. (3), the oscillations of both modes are characterized by the same spin gap and are degenerate at $H = 0$. The first equation in (3) describes well experimental points $\nu_{\perp 1}^{EA}$ with the parameters $\gamma_\perp =$ (2.65±0.02) GHz/kOe and $\nu_{\perp c}^{EA} = \gamma_\perp \sqrt{(2H_E + H_A)H_A}$ = (31.0±0.1) GHz. The theoretical dependence is shown by the solid line. The $\nu_{\perp c}^{EA}$ value matches the corresponding $\nu_{\|c}^{EA}$ value for the field orientation $H \| c$.

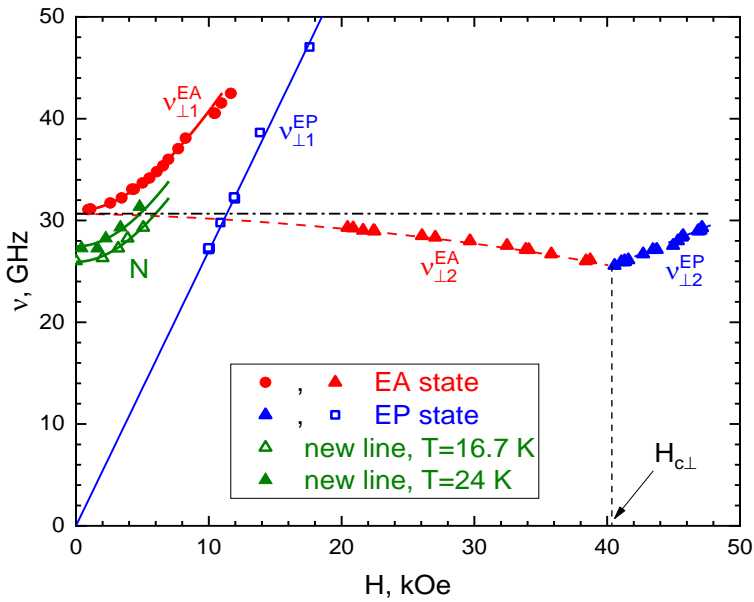
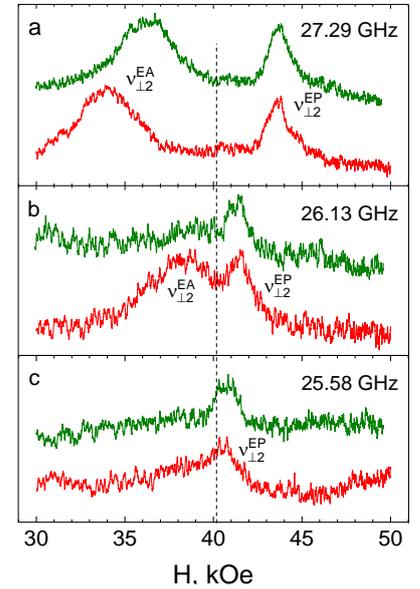

Fig. 6. AFMR frequency–field dependences in magnetic field $H \perp c$. The temperatures are $T = 4.2$ K for modes $\nu_{\perp 1}^{EA}$, $\nu_{\perp 2}^{EA}$, and $\nu_{\perp 2}^{EP}$; $T = 15.0$ K for $\nu_{\perp 1}^{EP}$; $T = 16.7$ K (open) and 24 K (closed) for mode N.

Fig. 7. Transformation of the AFMR spectrum of lines $\nu_{\perp 2}$ upon frequency variation. $T = 4.2$ K.

At the field orientation $H \perp c$, both a spontaneous EA–EP spin-reorientation upon temperature variation and a magnetic field-induced transition are also observed. In the EP state, the AFMR field–frequency dependences, neglecting the weak (according to the AFMR data) magnetic anisotropy in the basal plane, have the form [23]

$$\frac{\nu_{\perp 1}^{EP}}{\gamma_{\perp}} = H\sqrt{1 + \frac{|H_A|}{2H_E}} \approx H$$
$$\left(\frac{\nu_{\perp 2}^{EP}}{\gamma_{\perp}}\right)^2 = 2H_E|H_A| - \frac{|H_A|}{2H_E}H^2 \tag{4}$$

At $T = 4.2$ K, the oscillations of the mode $\nu_{\perp 1}^{EP}$ are not observed, since in the frequency range investigated in the experiment, the resonance field values are lower than the critical values required for the transition to the induced EP state. Figure 6 shows experimental points for mode $\nu_{\perp 1}^{EP}$ measured at temperature $T = 15.0$ K which is above the temperature of the spontaneous EA–EP transition.

The low-frequency oscillation branches below the spin gap in both the EA and EP states correspond to the second modes, $\nu_{\perp 2}^{EA}$ and $\nu_{\perp 2}^{EP}$, given by Eqs. (3) and (4). Under the condition $H_A \ll 2H_E$, the field dependences of $\nu_{\perp 2}^{EA}$ and $\nu_{\perp 2}^{EP}$ are identical for both states. In this regime, both frequencies exhibit a very weak dependence on the applied magnetic field, whose slope is determined by the ratio $H_A/2H_E$ for the EA state and $|H_A|/2H_E$ for the EP state. According to estimates for gadolinium ferroborate, this ratio does not exceed $\sim 10^{-4}$. In conventional antiferromagnets, the frequencies $\nu_{\perp 2}^{EA}$ and $\nu_{\perp 2}^{EP}$ soften significantly only as the field approaches $H \approx 2H_E$. On the scale of Fig. 6, the theoretical field dependences for both modes would thus appear as nearly horizontal lines, as indicated by the dash-and-dot line. Due to this weak dependence, resonant absorption for these modes is typically undetectable in weak magnetic fields using a field-sweep technique. Our experiment, in contrast, revealed a substantial field dependence for both $\nu_{\perp 2}^{EA}$ and $\nu_{\perp 2}^{EP}$. This result was initially reported in Ref. 13 for AFMR studies of $GdFe_3(BO_3)_4$ crystals containing a bismuth impurity. The observed substantial field dependence arises because the effective magnetic anisotropy, and consequently the spin gap, are themselves functions of the applied magnetic field.

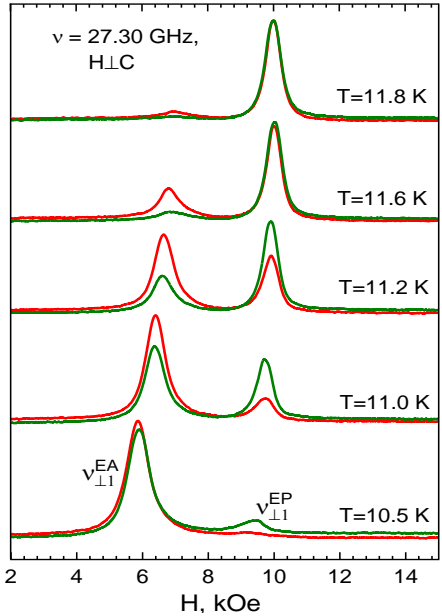
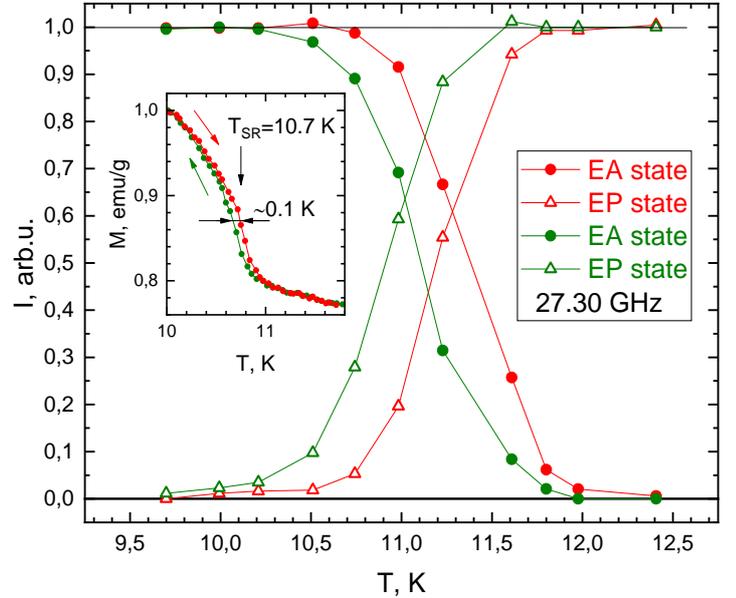

Fig. 8. Transformation of the AFMR spectrum lines $\nu_{\perp 1}$ in the region of the SR, $\nu = 27.30$ GHz, $H \perp c$. Red and green lines are recorded at the field-pulse leading and trailing edges, respectively.

Fig. 9. Temperature dependences of the AFMR line intensity for the EA and EP states during the SR measured at the leading (red dots) and trailing (green dots) edges of the magnetic field pulse. Inset: temperature hysteresis of the magnetization near the spin-reorientation, $H \perp c$.

The frequency-field dependence in this frequency region exhibits a characteristic V-shaped form: the resonance frequency first decreases almost linearly with the field, then reverses trend and increases linearly above a critical field $H_{c\perp}$. This kink is attributed to the field-induced EA–EP spin-reorientation transition. The measured value $H_{c\perp}$=40.5 kOe coincides with the critical field reported in Ref. 21 for T=4.2 K.

Figure 7 shows the evolution of the AFMR spectrum lines near the kink in the frequency-field dependence at T=4.2 K. In the EP state (Fig. 7a), the resonance fields for ascending (red) and descending (green) magnetic field sweeps coincide. In the EA state, however, they differ significantly. Furthermore, near the transition (Fig. 7b), the resonance peak in the EA state has an intensity comparable to that of the EP state when the field is increased, but becomes barely visible when the field is decreased. This behavior indicates that the field-induced EA–EP spin-reorientation for $H \perp c$ is accompanied by hysteretic phenomena, analogous to the case of $H \| c$.

The AFMR experiments for the $H \perp c$ orientation were conducted with the sample positioned in a waveguide to maximize the perpendicular component of the microwave magnetic field relative to the pulsed field ($\mathbf{h} \perp \mathbf{H}$). This geometry is optimal for exciting the $\nu_{\perp 1}$ resonance branches in both the EA and EP phases. Despite this alignment, a minor parallel component ($\mathbf{h} \| \mathbf{H}$) was present due to inherent inhomogeneities in the microwave field. This residual $\mathbf{h} \| \mathbf{H}$ component facilitated the detection of the $\nu_{\perp 2}$ branches, albeit with substantially lower intensity compared to the $\nu_{\perp 1}$ modes. Nevertheless, the signal-to-noise ratio for the $\nu_{\perp 2}$ branches was adequate for accurately determining their frequency-field dependencies.

Hysteretic phenomena are also observed for temperature variations at the $H \perp c$ field orientation. Near the spin-reorientation, the resonance line of the EA state (mode $\nu_{\perp 1}^{EA}$ in Fig. 6) vanishes, while the line characteristic of the EP state (mode $\nu_{\perp 1}^{EP}$) appears. Figure 8 illustrates the transformation of the AFMR spectrum across this transition. As the temperature increases, the intensity of the EA state's $\nu_{\perp 1}^{EA}$ mode decreases, while that of the EP state's $\nu_{\perp 1}^{EP}$ mode increases. The intensities recorded during increasing and decreasing magnetic field differ significantly. This thermal hysteresis is clearly demonstrated in Fig. 9, which plots the temperature dependence of the line intensities for both states. The hysteresis width at the transition is approximately 0.3 K, which is about three times larger than that of the magnetization measured by PPMS (see inset in Fig. 9). We attribute this discrepancy to the pulsed nature of the magnetic field used in our AFMR experiments.

The spin-reorientation temperature, estimated from the data in Figure 9, is approximately 11.2 K. This value is close to the $T_{SR}$ =10.7K determined from magnetic measurements on the same crystal (see inset, Fig. 9). The slight discrepancy is consistent with the estimated temperature measurement uncertainty of about 0.5 K in our AFMR experiment.

Figure 10 presents the temperature dependence of the AFMR fields for the GdFe$_3$(BO$_3$)$_4$ single crystal measured in a magnetic field $H \perp c$ at a frequency of 27.29 GHz (data are shown for the ascending field sweep). At low temperatures, only two resonance lines–denoted as modes $\nu_{\perp 2}^{EA}$ and $\nu_{\perp 2}^{EP}$ in Fig. 6–are observed at this frequency (see the AFMR spectrum in Fig. 7a). Given that the AFMR spin gaps are identical for the $H \| c$ and $H \perp c$ orientations, the temperature evolution of these resonance modes can be analyzed using the temperature-dependent behavior of the EA and EP gaps obtained for $H \| c$, shown in Fig. 5. The decrease in the resonance field for mode $\nu_{\perp 2}^{EP}$ with increasing temperature is attributed to the growth of the spin gap in the EP state, which continues up to ~20 K (Fig. 5). This drop in the resonance field is accompanied by a reduction in the mode's intensity, and the line is no longer observed above ~20 K.

The evolution of line $\nu_{\perp 2}^{EA}$, associated with the low-temperature EA phase, is more complex. Initially, the resonance field of this mode decreases as the spin gap narrows with rising temperature (Fig. 5). At $T \approx 6.5$ K, where the spin gap equals the measurement frequency, the system transitions to mode $\nu_{\perp 1}^{EA}$ (Fig. 6), and the resonance field begins to increase. This trend persists until the EA–EP spin-reorientation, which is marked by a crossover between frequency-field dependences $\nu_{\perp 1}^{EA}$ and $\nu_{\perp 1}^{EP}$ in Fig. 6. In the ensuing EP phase, the resonance field of mode $\nu_{\perp 1}^{EP}$ becomes temperature-independent, as predicted by Eq. (4) for this field orientation. Finally, at the Néel temperature $T_N$, mode $\nu_{\perp 1}^{EP}$ merges with the standard ESR line.

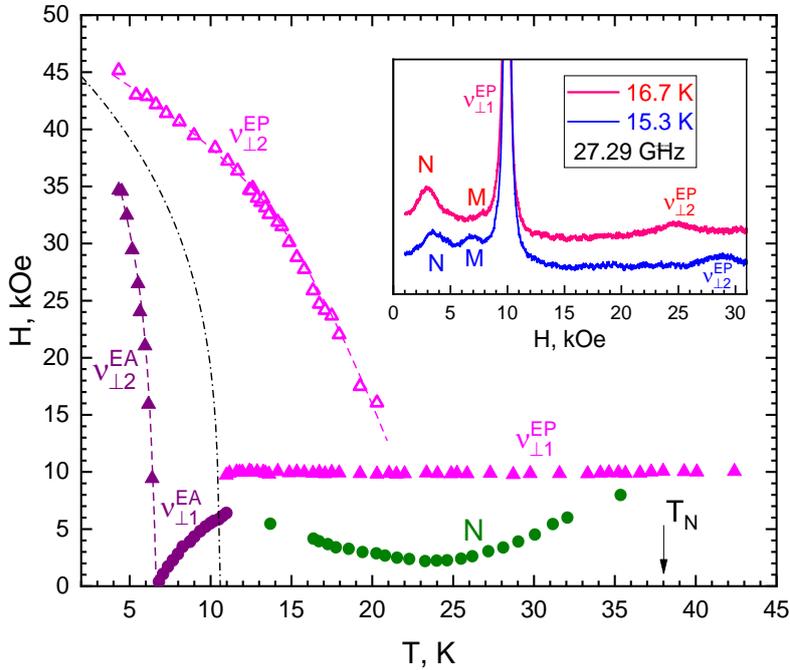

Fig. 10. Temperature dependences of AFMR resonance fields measured at 27.29 GHz in the magnetic field $\boldsymbol{H} \perp \boldsymbol{c}$. Dotted lines serve as an eye guide. The dash-and-dot line indicates the boundary between the EA and EP states from [21]. Inset: resonance spectra at a frequency of 27.29 GHz and temperatures of $T = $ 15.3 and 16.7 K.

Our analysis of the resonance field temperature dependences reveals an additional line (labeled N in Figs. 6 and 10) in the AFMR spectrum within the EP phase. Line N emerges 2–3 K above $T_{SR}$ and is distinct from the modes $\nu_{\perp 1}^{EP}$ or $\nu_{\perp 2}^{EP}$ described by Eqs. (4). At a fixed frequency of 27.29 GHz, its resonance field decreases with temperature, reaching a minimum near $T \approx 23$ K, before increasing again and converging with the ESR line at the Néel temperature. This non-monotonic behavior is attributed to the temperature dependence of the spin gap in the EP state, which exhibits a characteristic maximum around 23 K (Fig. 5). The correlation indicates that the frequency-field dependence of mode N is governed by the same spin gap as the other resonant branches. We will discuss the possible origin of this mode in Section 4.

### 4. Discussion
*4.1. First-order phase transition between the EA and EP states*

Using the temperature-dependent spin gaps in the AFMR spectrum of GdFe$_3$(BO$_3$)$_4$ (Fig. 5), we can calculate the effective magnetic anisotropy field and its change at the spin–reorientation transition based on the expressions for the gaps $\nu_{\parallel c}^{EA}$ and $\nu_{\parallel c}^{EP}$ in Eqs. (1) and (2). In the calculation, the exchange field $H_E^{Fe}$ for the iron subsystem is used, which has a value of $H_E^{Fe} = 700$ kOe at $T = 4.2$ K [12] and depends on the temperature as the Brillouin function $B_{5/2}$.

The spin–reorientation in GdFe$_3$(BO$_3$)$_4$ is a consequence of the competition between two opposing anisotropy contributions from the iron (Fe$^{3+}$) and gadolinium (Gd$^{3+}$) subsystems. In the isostructural YFe$_3$(BO$_3$)$_4$ crystal, in which Fe$^{3+}$ ions form a single magnetic subsystem [12], it was shown that the Fe$^{3+}$ anisotropy stabilizes EP state. In this material the anisotropy field is a smooth function of temperature which follows $B_{5/2}$ curve and reaches $-1.45$ kOe at 4.2 K [12].

In GdFe$_3$(BO$_3$)$_4$ the absolute values of anisotropy field is considerably smaller as can be seen from Fig. 11 and the curve exhibits a clear jump at $T_{SR}$ reversing its sign while maintaining its absolute value (Fig. 11). This jump occurs because the spin gap does not fully soften to zero at the transition as it is seen from Fig. 5. A complete softening would indicate a second-order transition with a continuous change in anisotropy. Instead, a residual gap $\nu_0 \approx 20$ GHz remains.

Theory (Eq. 1 and 2) shows that with a small anisotropy field ($H_A \ll 2H_E$), the expressions for the gaps $\nu_{\parallel c}^{EA}$ and $\nu_{\parallel c}^{EP}$ in the EA and EP states are identical except that the absolute value of $H_A$ is used in the EP state. Consequently, the incomplete softening forces $H_A$ to jump and change sign at $T_{SR}$. We note that the change of sign of the anisotropy field deduced from our analysis of AFMR data is in agreement with the magnetic phase diagram of GdFe$_3$(BO$_3$)$_4$ because the positive (negative) anisotropy field stabilizes EA (EP) states observed in the REXS measurements [11].

The non-monotonic behavior of anisotropy field is related to the anisotropic contribution of Gd subsystem. In Ref. [19] it was shown that the influence of R subsystem of Fe can be seen as an effective change of the second-order anisotropy constant. The effective anisotropy contains two contributions [19]:

$$K_{eff} = \frac{1}{2} K^{Fe} - \frac{1}{2}\left( \chi_{\parallel}^{R} \lambda_{\parallel}^{2} - \chi_{\perp}^{R} \lambda_{\perp}^{2} \right), \qquad (5)$$

where the second term is due to anisotropic contribution of R subsystem; $\chi_{\parallel,\perp}^{R}$ are components of the magnetic susceptibility tensor of R subsystem and $\lambda_{\parallel,\perp}$ are R – Fe exchange interaction constants. Under temperature decrease the second term in Eq. (5) becomes dominant causing the spin-reorientation to the EA state at $T = T_{SR}$. In Sec. 4.2 we will study this transition numerically and demonstrate that the 4-th order anisotropy constant $K_4$ is responsible for the incomplete softening of magnon mode and jump-like change of the anisotropy field at the spin-reorientation transition.

An external magnetic field can also induce the spin-reorientation transition. This is observed in GdFe$_3$(BO$_3$)$_4$ at $T < T_{SR}$ and in the field-dependent AFMR spectra at 4.2 K (See Fig. 6). As was mentioned above, the field dependences of the resonance frequencies $\nu_{\perp 2}^{EA}$ and $\nu_{\perp 2}^{EP}$ in the EA and EP states are determined by the ratio $H_A/2H_E$, which is no higher than $\sim 10^{-4}$. Therefore, for the field range of $H < 50$ kOe used in the experiments, the second terms, which take into account the dependence of the resonance frequency on the external magnetic field in Eqs. (3) and (4) can be neglected for the branches $\nu_{\perp 2}^{EA}$ and $\nu_{\perp 2}^{EP}$. Thus, frequency–field dependences for $\nu_{\perp 2}^{EA}$ and $\nu_{\perp 2}^{EP}$ in Fig. 6 are, in fact, the field dependences of the spin gap in the EA and EP states, respectively. Like the temperature dependence in Fig. 5, the field dependence of the gap has a kink in the critical field $H_{c\perp}$ of the spin-reorientation and the incomplete softening of the oscillatory mode with a residual gap of $\nu_{c0} \approx 25$ GHz. It is therefore not surprising that the field dependence of the effective anisotropy field (Fig. 12) also exhibits a jump at the spin-reorientation, while its absolute value remains unchanged.

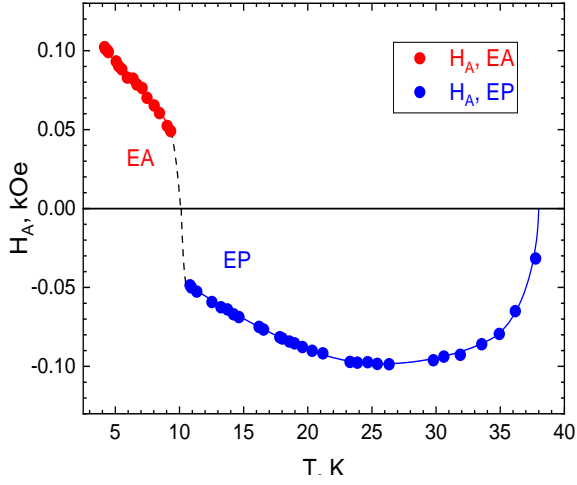 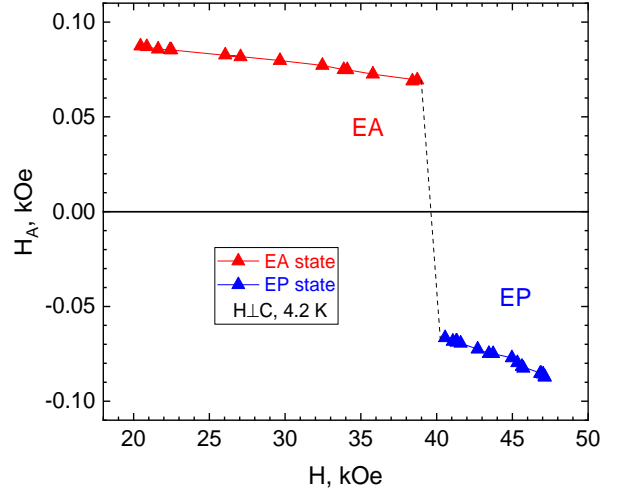

Fig. 11. Temperature dependence of effective magnetic anisotropy field $H_A$.

Fig. 12. Field dependences of the effective anisotropy field $H_A$.

### 4.2. Minimum microscopic model for the first-order phase transition

In this section, we introduce a minimum microscopic model to explain the first-order phase transition between the EA and EP states and the jump-like anisotropy change without a complete closure of the spin gap. Let us consider the effective Hamiltonian describing the magnetic behavior of the Fe subsystem:

$$H = H_{exc} + H_{anis}(T), \qquad (6)$$

where $H_{exc} = \sum_{i,j} J_{ij} S_i S_j$ and $H_{anis} = K_2(T) \sum_i \left(S_i^z\right)^2 + K_4 \sum_i \left(S_i^z\right)^4$

Here, the first term describes the dominant exchange interaction within the Fe subsystem, which stabilizes the AFM order. Because the microscopic details of the exchange interactions in the $GdFe_3(BO_3)_4$ are not reported in literature we construct the minimal model which satisfies two conditions: (i) the ground state reproduces the gross features of $GdFe_3(BO_3)_4$ magnetic structure, i.e. FM order within the ab-plane and AFM coupling along the c-axis; (ii) the calculated magnon excitations propagating along in-plane and out-of-plane directions reproduce the experimentally observed magnon bandwidths of Fe subsystem measured on isostructural compounds $NdFe_3(BO_3)_4$, $TbFe_3(BO_3)_4$ [24-26]. Thus, our model contains two exchange interactions: in-plane ferromagnetic (FM) $J_{ab} = -0.1$ meV and out-of-plane antiferromagnetic (AFM) $J_c = 1.2$ meV.

The second term in Hamiltonian (6) describes the *effective* magnetic anisotropy of the crystal and governs the spin reorientation transition. Note that microscopically, there are two distinct sources of the anisotropy: intrinsic anisotropy of Fe subsystem and additional term induced by interaction of Fe spins with anisotropic Gd subsystem. Equation (5) shows that both terms have similar functional form [19, 30] and therefore we cannot disentangle the two individual contributions. Instead, we consider their sum, which we call the *effective anisotropy*. In materials with the uniaxial anisotropy the quadratic term $K_2$ is usually much stronger than the quartic term $K_4$ and it determines the magnetic ground state. In turn, the effect of the quartic terms is especially pronounced in the proximity of phase transitions. For example, it was shown that the change of the sign of the effective anisotropy constant $K_2$ induces the G4–G2 spin-reorientation transitions in several rare-earth orthoferrites RFeO$_3$ [27-31], while the $K_4$ term ensures the occurrence of the spin-reorientation via two sequential second-order phase transitions with the inclined phase G4 + G2 stabilized at intermediate temperatures.

To describe the temperature-driven SR transition, we assume that, in the close proximity to $T_{SR}$, $K_2(T)$ can be expanded as

$$K_2(T) = K_0 + \alpha T , \qquad (7)$$

and therefore $K_2$ changes its sign at $T = -K_0/\alpha$, where $K_0 = 1.6 \cdot 10^{-3}$ meV and $\alpha = -8.41 \cdot 10^{-5}$ K$^{-1}$.

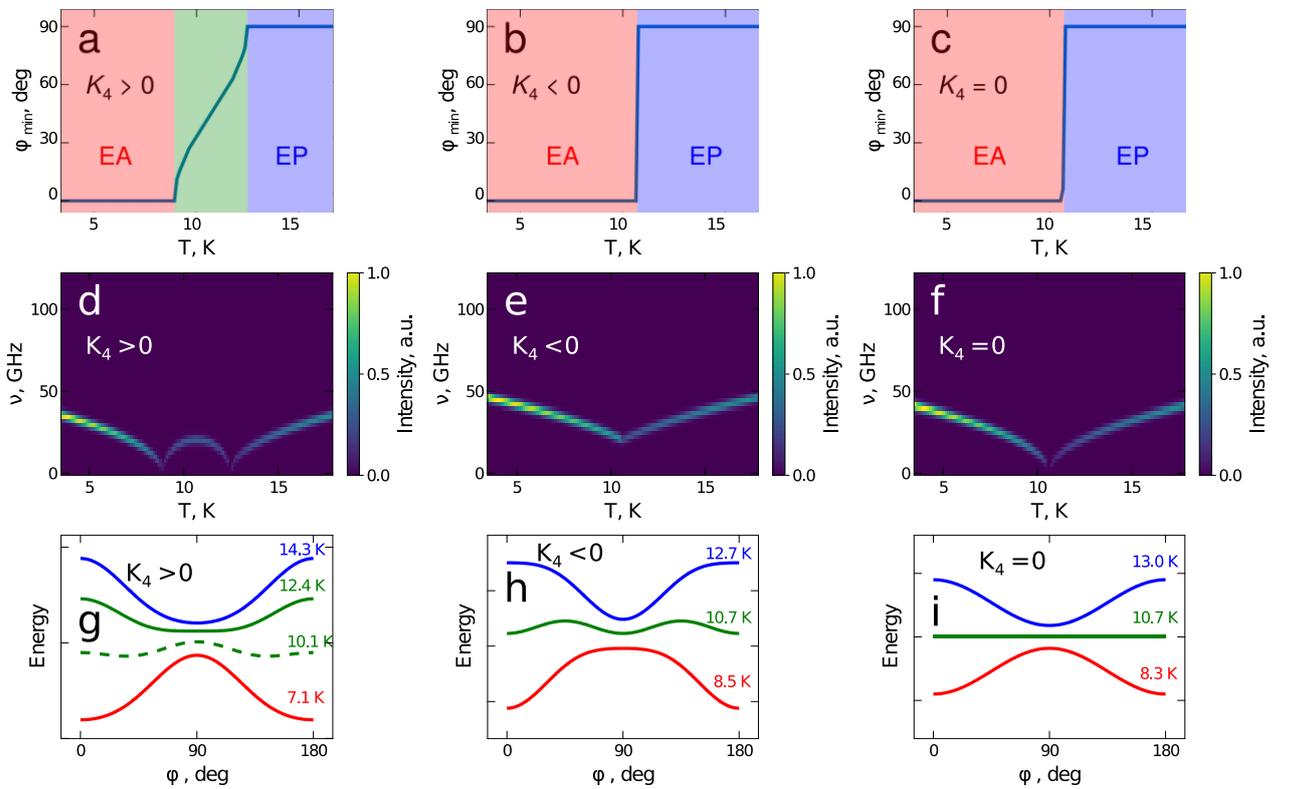

Fig. 13. Temperature dependences of (a–c) the Néel vector orientation and (d–f) the spin gap calculated for different $K_4$ values. (g–i) Angular part of the free energy calculated at different $K_4$ values and different temperatures indicated adjacent to the curves.

At the next step, we calculate the magnetic ground state and the Hamiltonian (6) spin dynamics using the Sunny software [32]. Figures 13a–13c show the temperature dependences of the angle $\varphi$ between the $c$ axis and the Néel vector calculated at different values of $K_4 = 0, \pm 0.0001$ meV, where 0 and $\pi/2$ correspond to the EA and EP states, respectively. Figures 13d–13f present the temperature dependences of the spin gap. In the absence of $K_4$, the temperature-induced SR occurs via one second-order phase transition with the complete softening of the magnon mode (Fig. 13f). At $K_4 > 0$, the transition splits into two second-order transitions and the inclined phase with a finite spin gap is formed between them (Fig.13a,d). Such a behavior was observed using the AFMR in YbFeO$_3$ compound [33]. In contrast to this, at $K_4 < 0$, the transition is abrupt and first-order, without complete magnon softening (Fig.13e), which is in qualitative agreement with our experimental data on the GdFe$_3$(BO$_3$)$_4$ single crystals shown in Fig. 5.

To get a physical insight why the type of a transition depends on the sign of the quartic anisotropy constant $K_4$, let us consider the angular dependence of the free energy at different states. Red and blue lines in Fig. 13i show the angular dependences of the free energy calculated within the EA and EP phases. Both curves follow the cosine-like function and have minima at π/2 (0) for the EP (EA) states. The spectra away from the transition temperature exhibit the gapped modes, because the gap is determined by the curvature ($\partial^2 F/\partial\varphi^2$) of the free energy near equilibrium[2] [34]. Now, let us consider the dynamics near the transition for three $K_4$ values. When $K_4 = 0$, the transition occurs at $K_2 = 0$, and the free-energy landscape is flat recovering the SU(2) symmetry of the exchange Hamiltonian (green line in Fig.13i). Therefore, the spectrum at $T_{SR}$ exhibits a gapless goldstone mode (Fig. 13f). At $K_4 > 0$, the curve calculated at the phase transition (green line in Fig.13g) is not flat but develops a broad shallow minimum with $\partial^2 F/\partial\varphi^2 = 0$ near the minimum, which explains the gapless magnon modes in Fig. 13d. The minimum continuously shifts with temperature increase between 0 and π/2, stabilizing the inclined phase. At $K_4 < 0$, the curve is ~cos(4φ) at the spin-reorientation transition (Fig.13h) and the energy minimum discontinuously switches from 0 to π/2, but the curvature near the minimum always remains finite, explaining the first-order nature of the phase transition and the presence of a finite gap in the spectrum (Fig. 13e).

### 4.1. *Origin of resonance line N*

When studying the temperature dependence of the resonance fields for the magnetic field orientation $\boldsymbol{H}\perp\boldsymbol{c}$, an additional line (line N) was observed in the AFMR spectrum (see inset in Fig. 10). This line does not correspond to either of the previously described oscillation modes ($\nu_{\perp 1}^{EP}$ or $\nu_{\perp 2}^{EP}$). This additional line was also observed in two other GdFe$_3$(BO$_3$)$_4$ single crystals from the same synthesis batch as the main crystals under study, as well as in a GdFe$_3$(BO$_3$)$_4$:Bi single crystal (~6% Bi$^{3+}$ impurity) that was previously investigated by resonance techniques [12, 13]. This reproducibility confirms that the additional line is not an artifact but a genuine magnetic response of the GdFe$_3$(BO$_3$)$_4$ system.

Figure 14 shows temperature dependences of the resonance fields for line 5 measured at different frequencies for the main GdFe$_3$(BO$_3$)$_4$ single crystal subjected to the resonance measurements. Similar dependences were obtained for other single crystals from the same synthesis batch, as well as for the GdFe$_3$(BO$_3$)$_4$:Bi single crystal.

The additional resonance line has the following features:
- it is only observed at the magnetic field orientation $\boldsymbol{H}\perp\boldsymbol{c}$;
- it appears in the EP state at temperatures several K higher than the temperature of the spontaneous spin-reorientation;
- the intensity of the resonance line is maximum at minimum resonance fields and rapidly decreases with increasing field.

---

[2] Calculations yield an additional gapless mode within the EP state which corresponds to the in-plane oscillations but here we focus on the second, gapped mode observed in AFMR

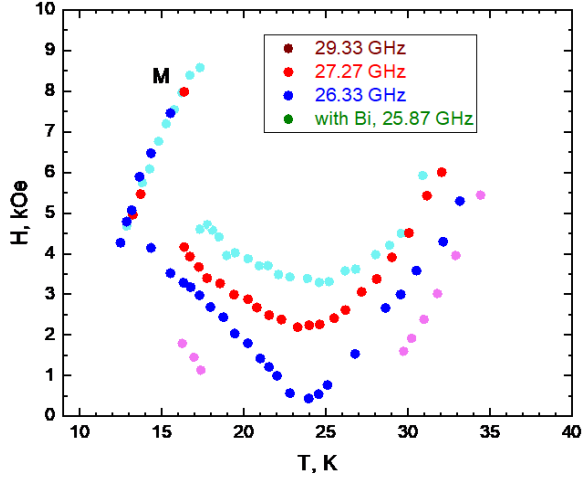
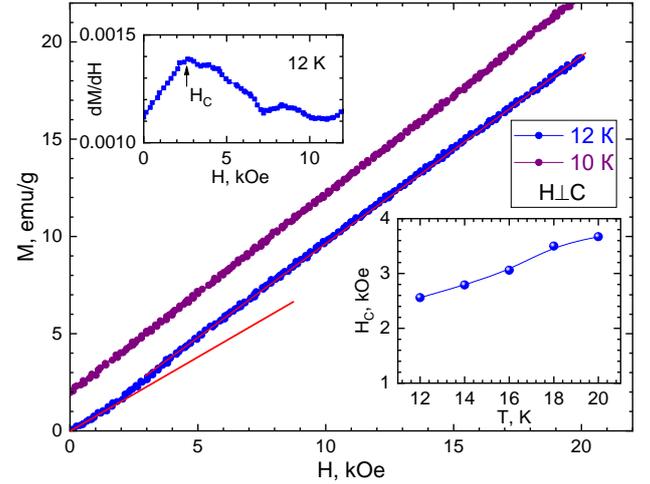

Fig. 14. Temperature dependences of the resonance fields for additional line N measured at different frequencies for the main GdFe$_3$(BO$_3$)$_4$ single crystal and the single crystal with the Bi impurity studied in [12, 13].

Fig. 15. Field dependences of the magnetization of the GdFe$_3$(BO$_3$)$_4$ in the magnetic field $H \perp c$ at temperatures of $T = 10$ K (shifted up by 2 emu/g) and 12 K. Inset: temperature dependence of critical field $H_C$.

The X-ray resonant magnetic scattering study of the GdFe$_3$(BO$_3$)$_4$ single crystal reported in [11] revealed a long-period incommensurate state with the wave vector structure (0, 0, 3/2±ε) and the spiral plane coinciding with the crystal basal plane. The incommensurability parameter ε depends on temperature and its maximum value does not exceed 0.002. The cited study showed that the incommensurate structure appears at the temperature $T_M$ which is 2 K higher than the temperature $T_{SR}$ of the spontaneous reorientation transition and exists up to the Néel temperature. An external magnetic field applied within the basal plane and exceeding a critical value destroys the incommensurate state, resulting in a commensurate state where the FIP phase emerges. In contrast, a magnetic field aligned with the trigonal axis, being perpendicular to the spiral plane, leaves the incommensurate state unaffected. These characteristics of the GdFe$_3$(BO$_3$)$_4$ magnetic phase diagram were also corroborated by measurements of the dielectric properties [17] and are illustrated in Fig. 1.

A comparison of the properties of the incommensurate state in GdFe$_3$(BO$_3$)$_4$ with the characteristics of the additional resonance line suggests that this resonance could originate from the spiral nature of the magnetic structure. This interpretation accounts for the specific temperature range over which the additional resonance mode exists. The absence of this mode for the $H \| c$ orientation is also consistent with this picture, as a magnetic field applied perpendicular to the easy plane (EP) does not alter the oscillation energy associated with the rotation of the antiferromagnetic vector within that plane. Consequently, for this field orientation, the spiral structure has no manifestation in the resonance spectrum. Finally, a magnetic field applied within the spiral plane distorts the spiral structure, transforming it first into a fan phase and then, with a further increase in field, into a commensurate structure. This evolution explains the observed decrease in the intensity of the resonance line with increasing magnetic field.

Furthermore, the long-period nature of the spiral structure allows it to be conceptually treated as a homogeneous EP antiferromagnetic state (the primary structural motif) with a small perturbation. This perspective provides a qualitative explanation for why the additional resonance line N is observed as a distinct feature against the background of the intense oscillation mode $v_{\perp 1}^{EP}$ characteristic of the homogeneous EP state.

The temperature-dependent resonance fields measured at various frequencies were used to reconstruct the frequency-field dependence for this mode. Representative results obtained at 16.7 K and 24 K are presented in Fig. 6. The experimental data are well described by the relation

$$\left(\frac{v}{\gamma}\right)^2 = \left(\frac{v_c}{\gamma}\right)^2 + H^2 \qquad (8)$$

with $\gamma = (2.73 \pm 0.09)$ GHz/kOe and spin gaps $v_c = (25.9 \pm 0.9)$ and $(26.5 \pm 0.9)$ GHz for T = 16.7 K and 24 K, respectively. These values of $v_c$ are consistent with the data obtained from Fig. 5 at the corresponding temperatures.

Fig. 14 also displays the temperature dependence of another feature in the AFMR spectrum, labeled M. This feature is frequency-independent; examples of AFMR spectra containing this anomaly are shown in the inset to Fig. 10 for two temperatures. Given its frequency-independent nature, feature M can be attributed to a non-resonant response of the microwave signal to a change in the magnetic state of the crystal. A similar non-resonant anomaly was observed at the EA-EP transition, as shown in Fig. 3c. We suggest that this anomaly is likely associated with the transition from the incommensurate state to the commensurate FIP phase. This transition and the field dependence of its critical temperature, $T_M$, were discussed in Refs. [11, 17]. Both studies confirmed that a magnetic field $\boldsymbol{H}\|c$, as expected, does not affect $T_M$, whereas for the $\boldsymbol{H}\perp c$ orientation, the critical field increases upon heating. However, the absolute values of the critical fields reported in these two studies [11, 17] differ by a factor of several.

To identify signatures of this transition in the static magnetic properties, we measured the magnetization of GdFe$_3$(BO$_3$)$_4$ single crystals in a magnetic field $\boldsymbol{H}\perp c$ at temperatures $T \geq 10$ K. At $T = 10$ K, the field dependence of the magnetization is smooth and featureless up to the maximum available fields (Fig. 15). At $T \geq 12$ K, a feature emerges in low fields, above which the magnetization increases linearly with the field up to the maximum measured field. A representative curve measured at $T = 12$ K is shown in Fig. 15, along with its derivative d$M$/d$H$ (left inset). The temperature dependence of the critical field for this anomaly is presented in the right inset. Given that no other features are observed in the magnetization curves, we associate this anomaly with the destruction of the incommensurate state in a magnetic field $\boldsymbol{H}\perp c$. The measured critical fields for the ICM–CM transition are consistent with those reported in [35] but are lower than the anomaly M determined from AFMR measurements. This discrepancy may be attributed to the pulsed nature of the magnetic field used in the AFMR spectrometer, which has a typical sweep rate of $\sim 10^7$ Oe/s. Despite these findings, the origin of the M anomaly in the AFMR spectrum remains an open question.

## 5. Conclusions

Our antiferromagnetic resonance (AFMR) study reveals key insights into spin-reorientation transition and magnetic anisotropy of the GdFe$_3$(BO$_3$)$_4$ single crystals. We find that the EA to EP spin-reorientation exhibits hysteresis in both temperature and magnetic field. This transition features an incomplete softening of an oscillation mode, causing a jump in the effective magnetic anisotropy field. Although the anisotropy field changes sign, its absolute value remains identical for the both states within the transition region.

Using a microscopic spin model, we successfully reproduced the spin-reorientation, including the incomplete softening of the spin gap. Our analysis demonstrates that the transition is driven by the temperature evolution of the effective anisotropy, arising from the distinct contributions of the Fe and Gd subsystems. The persistence of a finite spin gap is shown to be a direct consequence of a negative fourth-order effective anisotropy constant ($K_4 < 0$).

Additionally, we discovered a new AFMR oscillation mode under a magnetic field applied within the *ab* plane ($\boldsymbol{H} \perp \boldsymbol{c}$) at $T > T_{\mathrm{SR}}$. We argue that this mode is attributed to the dynamics of a long-period incommensurate magnetic structure, which is consistent with prior X-ray resonant magnetic scattering data. In the same temperature and field-orientation regime ($T > 12$ K, $\boldsymbol{H} \perp \boldsymbol{c}$), we observe an anomaly in the field dependence of the magnetization. We speculate that this anomaly corresponds to the suppression of the incommensurate state and a transition to a commensurate, high-field state.

### Acknowledgements

The magnetic properties of the crystals were investigated at the Krasnoyarsk Regional Center for Collective Use, Krasnoyarsk Scientific Center, Siberian Branch of the Russian Academy of Sciences.

### Funding

This study was carried out within the State Assignment of the Ministry of Science and Higher Education of the Russian Federation for the Kirensky Institute of Physics, Siberian Branch of the Russian Academy of Sciences.